\documentclass[aps,prl,twocolumn,superscriptaddress,showpacs,floatfix,amsmath,amssymb]{revtex4}

\usepackage{graphicx}% Include figure files
\usepackage{dcolumn}% Align table columns on decimal point
\usepackage{bm}% bold math
\usepackage{mathrsfs}
\usepackage{epsfig}
\newcommand{\fig}[1]{Fig.~\ref{#1}}
\newcommand{\be}[1]{\begin{equation}\label{#1}}
\newcommand{\ee}{\end{equation}}
\renewcommand{\vec}{\mathbf}

\begin{document}

\title{Recoil collisions as a portal to field assisted ionization at near-UV frequencies in the Strong Field Double Ionization of Helium}

\author{A. Emmanouilidou}
\affiliation{School of Physics, Georgia Institute of Technology, Atlanta, Georgia, 30332-0430}
 
\date{\today}
\begin{abstract}
We explore the dependence of the double ionization of the He atom on the frequency of a strong laser field while
keeping the ponderomotive energy constant. As we increase the frequency we find
that the remarkable ``finger-like" structure for high momenta recently found for $\omega=0.055$ a.u. \cite{Staudte, Rudenko} persists for higher frequencies. At the same time, at  $\omega=0.187$ a.u. a new X-shape structure emerges for small momenta that prevails in the correlated momenta distribution. The role of this structure as a signature of the frequency dependence of non-sequential double ionization is discussed. 

  \end{abstract}
\pacs{32.80.Rm, 31.90.+s, 32.80.Fb, 32.80.Wr }\maketitle   

The double ionization of the Helium atom driven by an infrared laser field
at intermediate intensities of $10^{13}-10^{15} W/cm^{2}$ has attracted considerable interest over the last few years as  a prototype
system for the study of the correlated emission of two-electrons in a driven atom. In this range of parameters double ionization proceeds via the rescattering mechanism \cite{Corkum}: the latter is a three-step process where first one electron tunnels to the continuum, then it is accelerated and finally is driven back by the laser field to its parent ion where it transfers energy and liberates the still bound electron.

Although the rescattering model has worked well in providing the interpretation of the basic strong field phenomena, such as ATI (Above Threshold Ionization), HHG (High-Order Harmonic Generation) and NSDI (Non-Sequential Double Ionization), recent refinements in experimental investigations \cite{Staudte,Rudenko}, have revealed additional structure in the latter, specifically the so-called ``finger-like" structure (V-shape) in the correlated momenta of the outgoing electrons; suggesting the presence of an underlying layer of effects. Their interpretation so far rests on a further interaction of the rescattering
electron with the nucleus, while in one version \cite{Rudenko}, the state of the core appears to play a
decisive role. At the same time, the possible influence of RESI (Recollision Excitation with Subsequent tunneling Ionization) in the finger-like structure seems to be ruled out according to ref. \cite{Staudte}. Moreover, work at 390 nm radiation, seems to suggest that the presence of the laser influences NSDI beyond the recollision \cite{Parker,Haan}.
 Thus, although considerable insight into the basic underlying mechanism for a finger-like structure
 in the momentum distribution of the electrons has been gained, it appears that a definitive quantitative interpretation may have to await further work.
 
 In the current letter, we explore the frequency dependence of NSDI. Much of its physical interpretation relies on the long wavelength ($\sim$ 800nm) under which essentially all of the experiments have been performed. Although it is understood that under much shorter wavelength, and comparable intensity, the rescattering mechanism eventually
 ceases to be valid, the transition from low to higher frequency remains an unexplored question.  Our aim
 is to explore the dependence, if any, of the finger-like structure on the wavelength of the radiation while keeping the ponderomotive energy constant. We show that the finger-like structure for large values of momenta recently found for $\omega=0.055$ a.u. persists for higher frequencies as well. At the same time a surprising X-shape like structure prevails for high frequencies. We find that this structure is related to a shift of the time of minimum electron-electron approach (recollision time) from $(2/3+n)T$ for small frequencies to T/2 for higher ones. In contrast to smaller frequencies, we find that for higher frequencies the target electron is significantly affected by the field and moves away from the nucleus before the rescattering electron reaches the nucleus.

Our approach is quasiclassical, but fully 3-dimensional. That alone would not be a sufficient justification, if it were not for the fact that it has proven quite useful in providing insight into problems of photon atom interactions \cite{Corkum1, Liu} for which fully quantum calculations entail prohibitive computational complexity. 
 At this time, no ab inito, fully 3-dimensional
 quantum calculation can cope with the computational demands it entails for the aspects addressed in this letter. Nevertheless, a number of judiciously chosen models \cite{reducedquantum1,reducedquantum2,reducedquantum3,Camillo}, including some classical, have proven quite useful in their interpretative and often predictive power. 

The quasiclassical model we use entails one electron tunneling through the field-lowered-Coulomb potential with a quantum tunneling rate given by the ADK formula \cite{ADK}. The longitudinal momentum is zero while the transverse one is given by a Gaussian distribution \cite{Liu}. The remaining electron is modeled by a microcanonical distribution \cite{Abrines}. For the evolution of the classical trajectories we use the full three-body Hamiltonian in the laser field, that is, $H=p_{1}^{2}/2+p_{2}^{2}/2
-Z/r_{1}-Z/r_{2}+1/|\vec{r}_{1}-\vec{r}_{2}|+(\vec{r}_{1}+\vec{r}_{2}) \cdot E(t)\hat{z}$, with E(t) the electric field (see \cite{Liu}) linearly polarized along the z-axis. The electric field
is a cos pulse that is on for 10 cycles and is then switched off in 3 cycles with a $cos^{2}$ envelope.
 We note however a difference between our method of propagation and the one used in \cite{Liu}: we employ regularized  coordinates \cite{regularized} (to account for the Coulomb singularity) which we believe result in a faster and more stable numerical propagation.

 To explore how the finger-like structure depends on the frequency of the radiation we explore the double ionization for three different frequencies 0.055 a.u, 0.11 a.u. and 0.187 a.u.
 In all three cases the ponderomotive energy $U_{p}=(E^{2}/(4\omega^2)$ is the same. Thus, the ratio of the time the electron needs to tunnel in the field-lowered Coulomb potential to the period of the laser field, the Keldysh parameter $\gamma=\sqrt{I_{p}/(2U_{p})}$ \cite{Keldysh}, is the same, where $I_{p}$ is the ionization potential of the He atom. For the frequencies under consideration, the respective 
 intensities $I$, with $I\propto E^2 $, are $3\times10^{14}$ W/cm$^2$,  $1.2\times10^{15}$ W/cm$^2$ and  $3.47\times10^{15}$ W/cm$^2$. In the following, we use the frequency to refer to each case.  For the calculations presented, at least $10^5$ double ionization events are obtained rendering our results quite accurate. Double ionizing trajectories are propagated even after the electric field is switched off until asymptotic values are reached.

% Our study unravels for the first time the mechanism for a transition to higher frequencies. The effect of the nucleus
 %becomes gradually more important, with the probability for the returning electron to backscatter from the nucleus accounting for almost half the
 %trajectories at $\omega=0.187$ a.u. compared to a 15\% at $\omega=0.055$ a.u. A transition to an intermediate regime takes place. The returning electron transfers energy to the bounded electron at times (2/3+n)T---the main times energy is transferred for $\omega=0.055$ a.u.---resulting in the usual double hump
% structure in the sum of the parallel momenta when the two electrons escape with a small inter-electronic angle. In addition,  a gradual shift
 %takes place of the main electron-electron encounter to a T/2 (maximum of the field) at $\omega=0.187$ a.u. resulting in an increased probabilty
 %for the two electrons to escape with smaller momenta and larger inter-electronic angle with a significant increase around 90$^{\circ}$. The interplay
 %of the two process with the effect of the nucleus becoming gradually dominant gives rise to a X-like structure in the correlated momenta.  

\begin{figure}
\scalebox{0.35}{\includegraphics{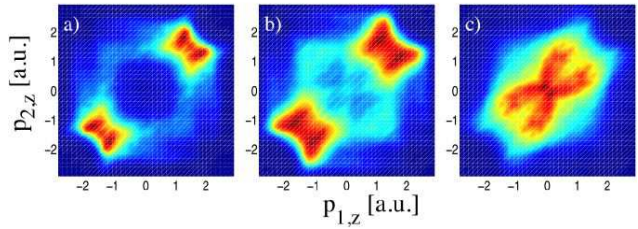}}
\caption{\label{momenta} Correlated momenta parallel to the field polarization for a) $\omega=0.055$ a.u. b) $\omega=0.11$ a.u. and c) $\omega=0.187$ a.u.}
\end{figure}

In \fig{momenta} we show the correlated momenta of the two electrons for the three different frequencies. A comparison of our result for $\omega=0.055$ a.u. with the experimental one for a pulse duration of 40fs, wavelength 800nm and peak intensity 4.5 $\times$ $10^{14}$ W/cm$^{2}$ \cite{Staudte} shows that we accurately capture the finger-like structure, which according to
ref. \cite{Staudte} is due to recoil collisions of the rescattering electron. Specifically, at $\omega=0.055$ a.u. this implies that the rescattering electron (denoted as electron 2) impacts the
other electron (denoted as electron 1) at times (2/3+n)T, with $n=0,1,2,...$ and T the period of the field, undergoing in addition a collision
with the nucleus resulting in its backscattering (recoil collision), with mostly reversing the direction of its velocity. The above times of recollision are also 
obtained in our calculation, through the examination of the average potential energy of the electron-electron interaction term and the identification of its maxima.

As a further check of our model, we show now that the finger-like structure we obtain (\fig{momenta}) is indeed due to recoil collisions. To this end, we identify the recollision time (the time of minimum approach of the two electrons) through the maximum in the electron pair potential energy. Further, we select those trajectories for which electron 2 backscatters from the nucleus, inverting the direction of its velocity. That is, $155^{\circ}<\vec{p}_{2,aft}\cdot \vec{p}_{2,bef}/|p_{2,aft}p_{2,bef}|<180^{\circ}$, with $\vec{p}_{2,bef/aft}$ the momentum of electron 2 just before and after the recollision time. The correlated momenta of the thus selected trajectories, as can be seen in Fig.2b, indeed account for the finger-like structure at $\omega=0.055$ a.u., also reported in ref \cite{Staudte}. In agreement with ref. \cite{Staudte} we find that this structure extends beyond the 2$\sqrt{U_{p}}$ maximum momentum limit (1.6 a.u. in our case).  Note first
that this structure persists for all three frequencies. In somewhat more details in Fig.2a we show the structure for correlated momenta with at least one of the two momenta having magnitude greater than $2\sqrt{U_{p}}$. We note that the trajectories shown in Fig.2b are a subset of those in Fig.2a and that for the remaining trajectories either 
electron 2 or electron 1 reverses its velocity but with a smaller recoil angle, that is, $90^{\circ}<\vec{p}_{i,aft}\cdot \vec{p}_{i,bef}/|p_{i,aft}p_{i,bef}|<150^{\circ}$.
Not evident in Fig.2a, we find that at the highest frequency 
the number of trajectories representing $p_{1}\vee p_{2}>2\sqrt{U_{p}}$ decreases and moreover the number of trajectories representing ``backscattering" in the sense of large recoil angle
also decreases; suggesting a reduction of recoil collisions. 
It is worth noting that we obtain the finger-like structure in Fig.2b for electrons
escaping asymptotically with a very small angle, almost parallel to each other.  To a smaller extent, the strong interaction with the nucleus also results in ``backscattering'' of either electron 2 or electron 1 with the two electrons escaping at a large angle, resulting in  related structure in the second and fourth quadrants of the correlated momenta.

Having established that the interaction of the rescattered electron with the nucleus is responsible for the finger-like structure, we discuss 
the imprint of the increasing frequency on the differential probabilties. We note that with increasing frequency the amplitude of excursion of the rescattering
electron diminishes. As the frequency changes from 0.055 a.u. to 0.187 a.u., we note the
following major changes: a) for increasing frequency the time of closest electron-electron approach shifts from $(2/3+n)T$ to T/2, when
the velocity of the rescattered electron due to the field is nearly zero; b) the examination of the average potential energy of electron 2 for the highest frequency reveals 
an increased effect of the nucleus.

\begin{figure}
\scalebox{0.35}{\includegraphics{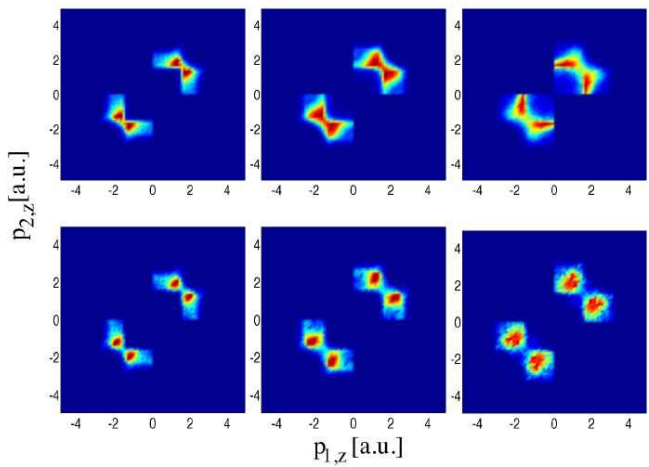}}
\caption{\label{recoil1} For frequencies from left to right  $\omega=0.055$ a.u., $\omega=0.11$ a.u. and $\omega=0.187$ a.u. we plot: a) (top panel) the correlated momenta using only the trajectories where 
$p_{1}\vee p_{2}>2\sqrt{U_{p}}$; b) (bottom panel) same as the top panel except that in addition electron 2 is backascattering with $155^{\circ}<\vec{p}_{2,aft}\cdot \vec{p}_{2,bef}/|p_{1}p_{2}|<180^{\circ}$. }
\end{figure}

\begin{figure}
\scalebox{0.5}{\includegraphics{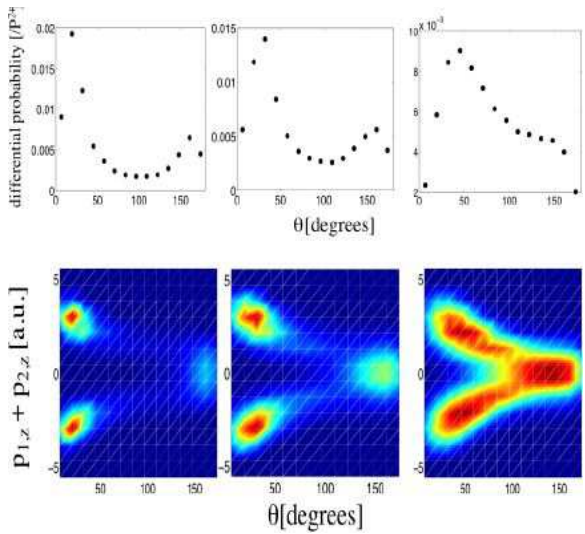}}
\caption{\label{recoil} For frequencies from left to right $\omega=0.055$ a.u., $\omega=0.11$ a.u. and $\omega=0.187$ a.u. we plot: a) (top panel) the distribution of the inter-electronic angles of escape binned in 14 intervals, 180$^{\circ}(l-1)/14<\theta<180^{\circ}/14\times l$ with l=1,...,14;
 b) (bottom panel) 
the sum of the parallel momenta as a function of the inter-electronic angle of escape.}
\end{figure}

The signatures of increasing frequency that appear to emerge are: 

a) a significantly less pronounced double hump in the parallel momentum distribution, see Fig.3b.
For a frequency of $\omega=0.055$ a.u. it is known that a less pronounced double hump structure results from an increased significance of the RESI mechanism versus the (e,2e) one \cite{Feuerstein,Weber}. For that
frequency in both mechanisms the main electron-electron encounters take place at a zero of the field, at times (2/3+n)T. However, while the release of the second electron happens at the same
time as rescattering for the (e,2e) process for the RESI it happens later at a maximum of the field. As a result, in the RESI process the electrons are released with smaller energy filling in the  ``valley'' between the two humps. Is it then possible that for the higher frequency RESI becomes more pronounced? On physical grounds, that might seem
reasonable since the ponderomotive energy responsible for the recollision excitation is the same while the photon energy is bigger. At this stage this is a conjecture that remains to be confirmed. 

b) While for the small frequency $\omega=0.055$ a.u. small inter-electronic angles of escape are favored, at $\omega=0.187$ a.u. this is no longer true, see Fig.3a. 
As a further check of the compatibility of our calculations with previous work \cite{Rudenko}, we have computed the inter-electronic angular distribution for
$\omega=0.055$ a.u. and $I=1\times 10^{15}$ W/cm$^2$ and find that a 180$^{\circ}$ escape is less probable compared to the $I=3\times 10^{14}$ W/cm$^{2}$ case; this is consistent with the fact that with increasing intensity---given that we remain within the non-sequential range---it is more likely that the second electron is ionized through an (e,2e) process. 
For the higher frequencies, already at $\omega=0.11$ a.u., it appears that inter-electronic angles of escape around  90$^{\circ}$ acquire more prominence; much more so for the
highest frequency currently considered as is evident in Figs.3a and b. Note that, already at $\omega=0.11$ a.u.
 while for small angles of escape
the electron-electron encounters take place at the same times as for $\omega=0.055$ a.u., for angles around 90$^{\circ}$ the encounters shift to times T/2. For the highest frequency T/2
is the most probable electron-electron encounter irrespective of the angle of escape.
Clearly, Fig.3b, even for larger frequencies when the two electrons escape almost parallel to each other we still find that the sum of the momenta components parallel to the field
is around its maximum possible value of 4$\sqrt{U_p}$. For increasing angles of escape the sum of the momenta components decreases significantly for the highest frequency.

Summarizing the results so far, we find that for $\omega=0.11$ a.u. the finger-like structure for momenta greater than 2$\sqrt{U_{p}}$ becomes more pronounced. However, for the frequency of 0.187 a.u., while the above structure is still present, somewhat unexpectedly a finger-like structure at smaller momenta emerges giving rise to an X-like pattern, see Fig.1. We have already discussed how we identify the trajectories where in addition to recollision, electron 2 backscatters from the nucleus giving rise to the finger-like structure for higher momenta. In a similar way, we identify the trajectories
where in addition to electron 2 undergoing a recollision, now it is electron 1 that backscatters from the nucleus for both electron momenta smaller than 2$\sqrt{U_{p}}$.  Using the latter trajectories we obtain the correlated momenta shown in Fig. 4. While for frequencies of 0.055 a.u. and 0.11 a.u. the trajectories with the additional feature of electron 1 ``backscattering'' from the nucleus merely complement the lower part of the the finger-like structure previously discussed for large momenta, this is not the case for $\omega=0.187$ a.u. For the highest frequency, these trajectories give rise to a V-shape or finger-like structure for small momenta in both the first and the third quadrant resulting in an overall X-shape structure that dominates the correlated momenta distribution, see Fig.1. Interestingly,
while for the case of the finger-like structure for large momenta the two electrons escape with a small angle, we find that for the new figure-like structure the electrons escape
with larger angles. Thus, it is the increased contribution of trajectories with angles of escape around 90$^{\circ}$ that are responsible for the prevailing X-shape structure for small momenta.

If one were to single out a major difference in behavior at the higher frequency, it is perhaps encapsulated in \fig{meanposition} which shows the relative position of the two electrons as a function of time.
In both Fig.5a and b we consider trajectories where in addition to the rescattering of electron 2, electron 1 recoils from the nucleus. In Fig.5a (small frequency) the position of electron 1 does not significantly change until electron 2 reaches the minimum distance from electron 1 which is practically the time of arrival at the nucleus. This happens around a zero of the field at $(2/3+n)T$.
 On the contrary, in Fig.5b ($\omega=0.187$ a.u. ) the time of minimum approach of the two electrons shifts to T/2 and while electron 2 is still approaching the nucleus,  
 electron 1 is already moving away from the nucleus. It is clearly seen that after the time of minimum electron-electron approach electron 1 responds both to the energy transfered
 from electron 2 but very importantly to the transfer of energy from the field.   For $\omega=0.055$ a.u. the transfer of energy to electron 1 takes place through the rescattering of electron 2. At high intensities---while in the non-sequential
 regime---this transfer of energy mainly takes place through an (e,2e) process while for smaller intensities trough an excitation and subsequent ionization from the field.  However, at high frequency the motion of electron 1 is significantly influenced by the field before the return of electron 2 close to the nucleus.

In conclusion, the prevailing X-shape structure we find for high frequencies is a new feature of NSDI that will motivate future experiments in this so far unexplored
regime of frequencies. Future theoretical work will focus on better understanding the interplay of an (e,2e) collision and the effect of the field and how the increased influence of the
field on the target electron before the approach of the rescattering electron to the nucleus is imprinted on the prevailing X-shape like structure.

 \begin{figure}[h]
\scalebox{0.35}{\includegraphics{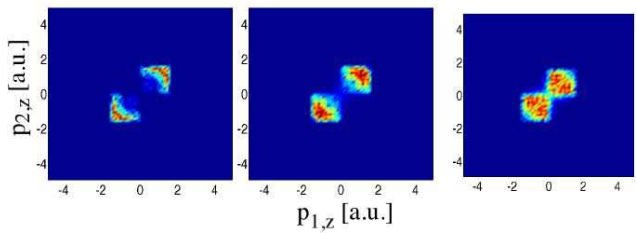}}
\caption{\label{90}For frequencies a) $\omega=0.055$ a.u., b) $\omega=0.11$ a.u. and c) $\omega=0.187$ a.u. we plot
the correlated momenta with electron 1 recoiling.}
\end{figure}

 \begin{figure}[h]
\scalebox{0.3}{\includegraphics{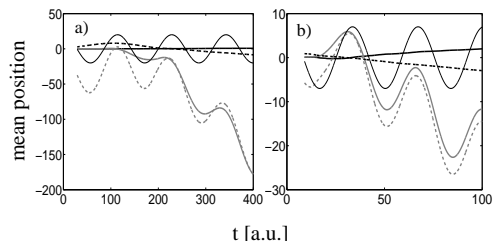}}
\caption{\label{meanposition}For frequencies a) $\omega=0.055$ a.u., b) $\omega=0.187$ a.u. we plot the mean position for electron 1 (solid lines) and electron 2 (dashed lines)
for the x component (black) and the z component (grey). Note that these averages are over the trajectories where electron 2 is ``born'' in the continuum in the negative
z direction, that is, the phase of the field when electron 2 tunnels is between $-\pi/2$ and $\pi/2$.}  
\end{figure}
\vspace{0.2cm}

Acknowledgment:  I am indebted to Peter Lambropoulos for suggesting the problem and for all the great discussions I had with him throughout this work. 
%I am also thankful for the hospitality of IESL-FORTH, Crete, where this work began.   

\end{document}